# A Worldwide Cost-based Design and Optimization of Tilted Bifacial Solar Farms

M. Tahir Patel[1], M. Ryyan Khan[2], Xingshu Sun[1], and Muhammad A. Alam[1]

[1]*Electrical and Computer Engineering Department, Purdue University, West Lafayette, IN 47907, USA*
[2]*Department of Electrical and Electronic Engineering, East West University, Dhaka, Bangladesh*

*Abstract –* **The steady decrease in the levelized cost of solar energy (LCOE) has made it increasingly cost-competitive against fossil fuels. The cost reduction is supported by a combination of material, device, and system innovations. To this end, bifacial solar farms are expected to decrease LCOE further by increasing the energy yield; but given the rapid pace of design/manufacturing innovations, a cost-inclusive optimization of bifacial solar farms has not been reported. In our worldwide study, we use a fundamentally new approach to decouple energy yield from cost considerations by parameterizing the LCOE formula in terms of "land-cost" and "module cost" to show that an interplay of these parameters defines the optimum design of bifacial farms. For ground-mounted solar panels, we observe that the panels must be oriented horizontally and packed densely for locations with high "land-cost", whereas the panels should be optimally tilted for places with high "module-cost". Compared to a monofacial farm, the modules in an optimized bifacial farm must be tilted ~15-20° higher and will reduce LCOE by ~8-10% in many locations of the world. The results in this paper will guide the deployment of LCOE-minimized ground-mounted tilted bifacial farms around the world.**

## 1. INTRODUCTION

Solar energy is spearheading the renewable energy growth around the world [1–4]. Levelized cost of energy (LCOE) is used as a metric to compare the economic viability of various sources of energy, e.g., wind, hydropower, coal, natural gas, solar energy etc.[5,6]. So far, solar energy still falls behind its competitors in this comparison [2,7]. There has been persistent and integrated efforts of research Institutes, PV industry, and government initiatives [4,5] to reduce LCOE of solar PV. Researchers are exploring new materials[8,9], the PV companies are optimizing manufacturing and module design [10], and farm-installers are developing new farm topologies (e.g., floating solar) and streamlining installation [11] [12].

Towards designing energy-efficient modules and farms [13]–[17], a new module design based on bifacial solar panels have shown ~50% increase in power output compared to monofacial panels [13]. The International Technology Roadmap for Photovoltaic (ITRPV) predicts that the worldwide market share for bifacial technology will increase to 40% by 2028 [14]. A recent literature review by Guerrero-Lemus et al. [15] explains the optimism surrounding the initiation, growth, and scalability of this technology. To this end, various standalone module designs have been recently investigated numerically [16–20] and experimentally [21–24]. These demonstrate the dependence of irradiance intensity, spectral distribution, the fraction of direct, diffuse, and albedo light, etc. on the design of stand-alone bifacial modules. Guo et al. [25] have done a global analysis of east-west-facing vertical bifacial modules, Ito et al. [26] have presented a comparison of vertical bifacial modules and tilted modules, and Sun et al. [19] have provided a global perspective of bifacial modules with optimized tilt angle, azimuth angle, and elevation. The bifacial gain of stand-alone bifacial modules is significant enough to support the optimistic view of the technology.

Unfortunately, a combined effect of panel-to-panel (mutual) and panel-to-ground (self) shading in a solar farm may erode the perceived advantage of stand-alone bifacial farms. Recently, Appelbaum [27] has investigated the effect of shading when bifacial modules in a farm are installed in multiple rows for east-west and south-north orientations. Khan et al. [28] have done a global analysis of vertical bifacial solar farms and observed a 10-20% energy gain for practical row-spacing. In a more recent study [29], they have proposed a ground-sculpting of farms to enhance the power output and achieve ~50% bifacial gain. Moreover, the vertical farm design reduces soiling and cleaning cost. In fact, limited experimental study in Tucson (Arizona, USA) and Forst (Lausitz, Germany) do show the vertical bifacial farms outperforming the optimally tilted bifacial farms for certain months/weather conditions [30]. It is not clear if these conclusions can be generalized to all locations in the world.

Most importantly, the reports to date focus on bifacial energy gain, but it is unclear if the gain is sufficient to offset the additional cost of a bifacial solar farm to make the technology commercially viable. A recent study [31] has offered a simplified cost-based analysis of bifacial farms: the analysis focused only on module cost (land cost was ignored) and a collection of stand-alone modules (self and mutual-shading were ignored). In other words, the cost was underestimated, while energy yield was overestimated. A generalization of the previous study that predicts the LCOE-optimized bifacial farm design (including land cost and mutual shading) would be of great interest.

The calculation of farm-level levelized cost of energy (LCOE) is difficult because some of the factors (e.g. land cost, module and installation cost, degradation rates, and bankability) may not be known for years to come. Therefore, we need to reformulate/re-parameterize the LCOE calculation in a way that allows design optimization and decision making even when all the cost details are not known. The new approach will have an additional benefit of providing a clear understanding of the impact of various parameters on bifacial LCOE and a worldwide mapping/analysis will show us a trend in the economic viability of the bifacial solar cell technology.

In this paper, we explore the design optimization of tilted bifacial solar farms for all the locations across the globe. The objective is to find the key design parameters, appropriately model the insolation and light collection at the farm-level, and subsequently incorporate a parameterized cost analysis to eventually find an optimized design that minimizes the cost of electricity produced. Our generalized model shown in Fig. 1 will reproduce, as special cases, various types of farms designed to date. For example, specialized designs are achieved by fixing one of the seven bifacial farm parameters defined in Table 1, namely, $\beta = 90°$ defines a vertical bifacial farm; $p = \infty$ yields a stand-alone bifacial PV module; and turning off the back-surface light-collection creates a monofacial solar farm. Other designs can be obtained by varying several parameters simultaneously.

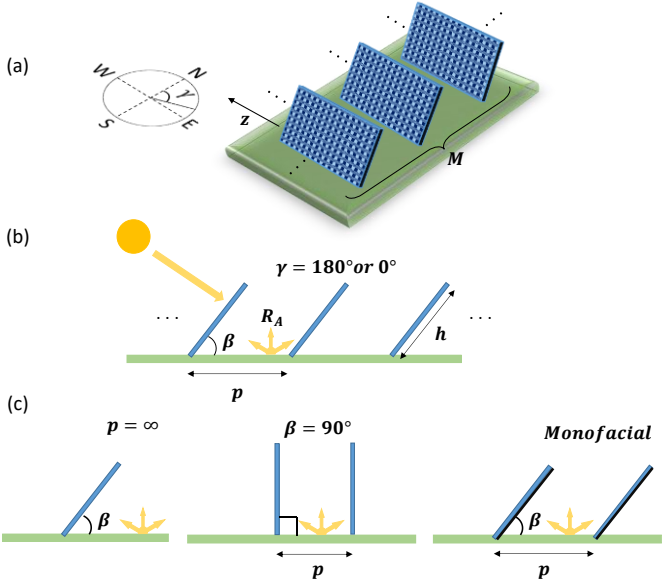

Fig. 1 (a) A 3-D Schematic of a generalized tilted bifacial solar farm. (b) A 2-D view of arrays of ground-mounted panels, assuming infinite (or unit) length in the third dimension. (c) Derived designs, namely, standalone module, vertical farm, and monofacial farm, as derivatives of the general design.

Table 1: Physical and economic parameters of a ground-mounted tilted bifacial farm

| # | Parameters | Definition |
|---|---|---|
| 1. | Pitch ($p$) | Row-to-row distance between the bottom edges of consecutive arrays |
| 2. | Height ($h$) | Height of the panels |
| 3. | Array tilt angle ($\beta$) | Angle between the array (panels) and the ground (horizontal) |
| 4. | Array azimuth angle ($\gamma$) | Angle between the projection of normal to the front face of the |

arrays on the ground and the North Pole

| 5. | Albedo ($R_A$) | Fraction of incident light reflected from the ground |
| 6. | Land cost ($c_L$) | The cost associated with a unit area of land covered by the panel arrays in the solar farm |
| 7. | Module cost ($c_M$) | The cost of a single solar module per unit height (includes fabrication and installation costs) |

In Sec. 2 we explain the models used in our study – In Sec. 2.1, we start with re-parameterization of LCOE in terms of land cost ($C_L$), module cost ($C_M$), and yearly energy yield ($YY$). Next, in Sec. 2.2, we describe the methodology to calculate the local irradiance by using the available meteorological data. The outcome of this analysis will provide the illumination/energy incident on the solar farm area. We then discuss, in Sec. 2.3, the light collection model for estimating the fraction of incident light absorbed by front/rear faces of the solar modules to calculate the power output from the farm in terms of annual energy yield. Thenceforth, in Sec. 3, we optimize the bifacial farm to minimize the levelized cost of energy. Finally, in Sec. 4, we present the results of the simulations and a global perspective of minimized levelized cost of energy for optimum system parameters. Our conclusions are summarized in Sec. 5 of the paper.

## 2. MODELING

An LCOE-aware modeling and optimization of bifacial solar farm involves a series of calculations. In this section, we will focus on three aspects of the modeling framework: (A) Re-parameterizing LCOE, (B) Irradiance modeling, and (C) Collection of light. These topics are discussed below.

### 2.1. Levelized Cost of Energy (LCOE)

LCOE is defined as the ratio of the total cost of a PV system and the total energy yield of the system over its lifetime [32], i.e.,

$$LCOE = \frac{\text{Total Cost (\$)}}{\text{Total Energy Yield (kWh)}}$$

$$= \frac{C_{sys}(Y=0) + (\sum_{k=1}^{Y} C_{om}(k)) - C_{rv}(Y)}{E(Y)}$$

(1)

where $C_{sys}(Y=0)$ is the fixed cost paid once during installation (i.e., $Y = 0$) of the system. This includes the cost of modules ($c_{m,0}$), the cost of land ($c_{l,0}$), and the balance of system cost ($c_{bos,0}$) such as labor, permit, racks, inverters, etc. The recurring operations and maintenance cost ($C_{om}$) scales with the cost of maintaining individual modules ($c_{om,m}$) and the cost of maintaining the land ($c_{om,l}$). Finally, $C_{rv}$ is the residual value of the modules ($c_{rv,m}$), the land ($c_{rv,l}$), and the equipment to be regained when the farm is decommissioned. $C_{om}$ and $C_{rv}$ are a function of the lifetime (number of years, $Y$) for which the solar farm is operated.

The total cost (numerator) in the LCOE expression in Eq. (1) can be *equivalently* written as the sum of effective module cost ($C_M$), effective land cost ($C_L$), and fixed balance of systems cost ($C_{bos,f}$), as shown in the following equation.

$$LCOE \equiv \frac{C_M + C_L + C_{bos,f}}{E(Y)} \qquad (2)$$

where

$$
C_M = c_{m,0} + \left( \sum_{k=1}^{Y} \left( c_{om,m}(k)(1+r)^{-k} \right) \right)
$$
$$
- C_{rv,m}(Y)(1+r)^{-Y}
$$
$$
C_L = c_{l,0} + \left( \sum_{k=1}^{Y} \left( c_{om,l}(k)(1+r)^{-k} \right) \right)
$$
$$
- C_{rv,l}(Y)(1+r)^{-Y}
\qquad (3)
$$

Subscripts $m, l,$ and $bos$ stand for module, land, and balance of systems. Moreover, $r$ is the discount rate that normalizes future costs in terms of current cost in order to have a fair comparison using a single metric.

The key insight of Eq. (2) is that the costs associated with a solar farm reflects two essential costs, namely, effective module cost ($C_M$) and effective land cost ($C_L$). The costs that scale with the number of modules (e.g., module size, material cost, racking, wiring in panels, inverters etc.) are included in $C_M$. Those costs that vary with the area of land (e.g., cost of land, fencing, cost for land curing/sculpting, etc.) are collected in $C_L$. Since the (typically small) fixed cost associated with balance of systems ( e.g. permit cost) does not scale with the number of modules and land, hence, we collect them into the residual cost, $C_{bos,f}$.

The denominator of Eq. (2) describes the total energy yield $E(Y)$ of the solar farm:

$$E(Y) = \sum_{k=1}^{Y} E_0 \,(1-d)^{-k}(1+r)^{-k} \qquad (4)$$

Here, $E_0 = P_0 \times T_Y$ is the energy output for the first year expressed as the product of first year power output ($P_0$) and total number of active hours in a year ($T_Y$). The yearly energy degradation rate ($d$) defines the lifetime ($Y$, in years) of the solar farm. The discount rate ($r$) accounts for the following fact: If we continue to sell a unit of energy for c \$/watt, the present value of the lifetime revenue must account for the fact that future earnings are less valuable than present earning.

Given the dimensions of a solar farm and the modules installed, LCOE expression in Eq. (2) can be simplified even further, i.e., :

$$LCOE = \frac{\mathbb{C}_M(r).h.M.Z + \mathbb{C}_L(r).p.M.Z + C_{bos,f}}{YY(p,h,\beta,\gamma,R_A).M.Z.h.\chi(d,r)} \qquad (5)$$

where $\mathbb{C}_M$ is the cost per unit meter of module (height), $\mathbb{C}_L$ is the cost per unit meter of land (pitch), $M$ is the number of rows/arrays of modules and $Z$ is the number of modules in an array (in the z-direction, into the page). $YY(= E_0)$ is the yearly energy yield per meter of a pristine module for one period/pitch ($p$) such that the yearly energy of the farm is $E(Y) = YY.M.Z.h.\chi(d,r)$, where $\chi = \sum_0^Y (1-d)^k (1+r)^{-k}$. $YY$ is a function of the physical design parameters ($p,h,\beta,\gamma,R_A$). Further, the residual cost associated with the balance of system ($c_{bos,f}$) is typically negligible as compared to the essential costs, $C_M$ and $C_L$ [12], and does not affect the design optimization of the farm. With these considerations, we arrive at the 'essential levelized cost of energy' ($LCOE^*$), as follows:

$$LCOE^* \equiv \frac{LCOE.\chi}{\mathbb{C}_L} = \frac{\mathbb{C}_M/\mathbb{C}_L + p/h}{YY(p,h,\beta,\gamma,R_A)} = \frac{p/h + M_L}{YY} \qquad (6)$$

Eq. (6) defines an important design parameter $M_L(\equiv \mathbb{C}_M/\mathbb{C}_L)$ as the ratio of cost of module per unit length (height) over cost of land per unit length (pitch). Note that Eq. (6) expresses $LCOE$ explicitly in terms of the seven fundamental farm design variables shown in Table 1.

The parametric reformulation of $LCOE$ in terms of $LCOE^*$ decouples the cost analysis (embedded in $M_L$) from design considerations (reflected in $p/h$ and $YY$). In other words, we can optimize a farm with $M_L$ as a parameter, realizing that $M_L$ will evolve as the process technology evolves. Equation (6) shows that $LCOE$ is proportional to $LCOE^*$, since $\mathbb{C}_L$ and $\chi$ are location-specific constants. Therefore, minimizing $LCOE$ is equivalent to minimizing $LCOE^*$ for a given location. In the following sections, we will estimate the yearly energy yield at a particular location and then account for the costs to finally arrive at the location-based $LCOE^*$.

## 2.2. Irradiance model for calculating $I_{DNI}$ and $I_{DHI}$

To estimate the energy yield of a solar farm, the simulation proceeds in three steps. First, we calculate the amount of sunlight incident at a particular location defined by its latitude and longitude. Next, we quantify the amount of light collected by the solar panels installed at that location. Finally, we find the daily and yearly power/energy-output of the farm.

### 2.2.1. Location-based light intensity

To calculate the temporal solar irradiance data at a particular location, we need to know the solar trajectory (or Solar path) and the intensity of light [19,28]. The solar path can be acquired/simulated by using the NREL's solar position algorithm [33] which has been implemented in Sandia PV modeling library (PVLib) [34]. This gives us the zenith angle ($\theta_z$) and azimuth angle ($A$) of the sun at any time of the day on a given date for the desired location. Here, $\theta_z$ is the refraction-corrected zenith angle, which depends on altitude and ambient temperature. We use the Haurwitz clear sky model to generate the Global Horizontal Irradiance (GHI or $I_{GHI}$) [35,36] with a time resolution of one minute. The clear sky model, however, often overestimates insolation, especially when the atmosphere is cloudy or overcast. Hence, to account for local variation of GHI caused by cloudiness and altitude, we scale the integrated

GHI over time to match the satellite-derived 22-year monthly average GHI data from the NASA Surface meteorology and Solar Energy database [37]. Therefore, our modeling framework takes into consideration the impacts of geographic and climatic factors to model the location-specific solar irradiance.

### 2.2.2. Direct and Diffused light from GHI

Since GHI is measured on a flat ground while the solar panels are tilted, we need to decompose the amount of direct light, called Direct Normal Irradiance (DNI or $I_{DNI}$), and the amount of diffused (scattered) light, called the Diffuse Horizontal Irradiance (DHI or $I_{DHI}$), which are related to the GHI as follows:

$$I_{GHI} = I_{DNI} \times \cos(\theta_Z) + I_{DHI} \qquad (7)$$

Since we have one equation and two unknown variables, $I_{DNI}$ and $I_{DHI}$, thus, we estimate one of the variables, $I_{DHI}$, using the Orgill and Hollands model which empirically calculates the diffuse fraction using the clearness index of the sky, defined as the ratio between $I_{GHI}$ and extraterrestrial irradiance ($I_0$) on a horizontal surface, as shown below.

$$k_T = \frac{I_{GHI}}{I_0 \cos \theta_Z} \qquad (8)$$

Here, $I_0$ comes from an analytical expression in Ref. [38].

Once we calculate $I_{DHI}$ using $I_{GHI}$ and $k_T$, Eq. (7) gives us $I_{DNI}$. Further, instead of the isotropic sky model [39], we deploy the Perez model [38,40,41] to account for the anisotropic (angle-dependent) decomposition of $I_{DHI}$. This fixes the overestimation of energy yield due to isotropic model [23]. A similar calculation for estimation and decomposition of solar insolation has been previously done by others [19,28].

### 2.2.3. Angle of incidence (AOI)

We need to calculate the AOI to find the component of direct light ($I_{DNI}$) falling on the tilted panel's front and/or the back face (depending on the tilt angle). The AOI for N-S facing tilted panels can be analytically expressed as [42]:

$$AOI_{Front} = \theta_F = \cos^{-1}\Big(\cos\theta_Z \cos\beta \; + \\ \big(\sin\theta_Z \sin\beta \cos\big((\gamma - 180) - (A - 180)\big)\big)\Big) \qquad (9)$$

$$AOI_{Back} = \theta_B = \cos^{-1}\Big(\cos\theta_Z \; \cos(180 - \beta) \; + \\ \big(\sin\theta_Z \sin(180 - \beta) \cos\big((\gamma - 180) - A\big)\big)\Big) \qquad (10)$$

where $\theta_F$ and $\theta_B$ are angles of incidence on front and back surfaces of the panel, respectively, and $A$ is the azimuth angle of the sun.

We now have the irradiance and angle of incidence information to proceed with estimation of light collection and energy generation by the solar panels.

### 2.3. Collection of light for calculating $YY(p, h, \beta, \gamma, R_A)$

The panels have height $h$, tilted at an angle $\beta$, separated by pitch (or period) $p$, and are oriented at array azimuth angle $\gamma = 180°$ (i.e., south-facing panels) for farms in the northern hemisphere and $\gamma = 0°$ (i.e., north-facing panels) for farms in the southern hemisphere (see Fig. 1). For simplicity, we assume that the arrays run sufficiently long in East-West direction that the edge effects can be neglected. The collection of light on panels requires different approaches for the three components of irradiance i.e., direct, diffuse and albedo light, and hence, we will analyze them individually. We will first calculate the light collection and energy output for each panel/array and then account for the energies from all the arrays to estimate the total energy yield from the farm. Our model for light collection is similar to and based on the one developed by Khan et al. [28]. Under direct and diffuse irradiance, we assume $\eta_{dir} = 18.9\%$ and $\eta_{diff} = 15.67\%$ [28,43–45], respectively. Moreover, the cell efficiency for front and the back faces are experimentally found to differ by 1–2% [15]. For simplicity, we neglect this difference. It is important to note that there can be partial shading of panels or non-uniform illumination during the day. This can cause some of the solar cells in a panel to reach reverse breakdown. This is mitigated by placing bypass diodes (3 in our case) connected across different sub-sections of series-string. The effect of shading and bypass diodes on the power output of the panel is discussed by Deline et al. [46]. Our model accounts for partial shading effects [28]. Now, let us look at the three individual components of incident light.

### 2.3.1. Direct light collection

Starting from $I_{DNI}$ estimated earlier, we find that the component of direct illumination normal to the front surface of the panel is given by $I_{DNI} \cos \theta_F$. Next, we use an empirical model [39R] to incorporate the angle-dependent reflectivity ($R(\theta_F)$) of the panel. Finally, including the efficiency $\eta_{dir}$, we arrive at the power output per unit panel area.

$$I_{PV:DNI}^{F,Panel} = I_{DNI} \cos\theta_F \left(1 - R(\theta_F)\right)\eta_{dir}; \; l > shadow \\ = 0; \qquad\qquad\qquad l \leq shadow \qquad (11)$$

The contribution of power from the shaded area of the panel is assumed to be zero. Thus, the power output per unit height of a panel is equal to $I_{PV:DNI}^{F,Panel}$. Now, the power output per pitch of the farm ($I_{PV:DNI}^{F,Farm}$) is given by:

$$I_{PV:DNI}^{F,Farm} = I_{DNI} \cos\theta_F \left(1 - R(\theta_F)\right)\eta_{dir} \qquad (12)$$

Similarly, we can find the direct light collection for back surface ($I_{PV:DNI}^{B,Farm}$) with the tilt angle equal to $180° - \beta$ and $AOI_{Back} = \theta_B$. Therefore, the total power output per unit farm area due to direct light ($I_{PV:DNI}^{Farm}$) is the sum of $I_{PV:DNI}^{F,Farm}$ and $I_{PV:DNI}^{B,Farm}$.

### 2.3.2. Diffuse light collection

The estimation of the diffuse light collection is more involved as compared to direct light. Using the widely-accepted technique of the average diffuse masking angle [47] overestimates the magnitude of collected diffuse light, especially for panels with large tilt angles. A view factor approach has been previously applied to find the average diffuse light collection [48]. Here we find diffuse light incident on each point on the panel to appropriately find the incident diffuse light distribution over the panel faces. This model ensures a more accurate representation of the effect of non-uniform illumination (from direct, diffuse, and albedo light).

The spacing between the arrays is such that there is no mutual shading for direct light for most part of the day, however, the isotropic diffuse light does result in mutual shading. The diffuse light falling on the front face of a panel is partially shaded by the panels of the adjacent arrays. The point at length $l$ from the bottom of the panel views an angle $\psi(l)$ shaded or masked by the adjacent panel (see Fig. 2). The angle $\psi(l)$ is geometrically calculated as:

$$\psi(l) = \tan^{-1}\left[\frac{(1 - l/h)\sin\beta}{p/h - (1 - l/h)\cos\beta}\right] \quad (13)$$

The diffuse light intensity on the front face, at position $l$ along the panel is given by $I_{DHI} \times F_{dl-sky}(l)$. The view factor from $dl$ to the unobstructed sky, $F_{dl-sky}(l) = 1/2(1 + \cos(\psi(l) + \beta))$ [49]. Thus, we arrive at the power generated per unit panel area.

$$\begin{aligned} I_{PV:DHI}^{F,Panel}(l) &= I_{DHI} \times F_{dl-sky}(l) \times \eta_{diff} \\ &= I_{DHI} \times \frac{1}{2}(1 + \cos(\psi(l) + \beta))\,\eta_{diff} \end{aligned} \quad (14)$$

Hence, the total integrated power output per pitch is given by

$$I_{PV:DHI}^{F,Farm} = I_{DHI}\,\eta_{diff}/h \int_0^h \frac{1}{2}(1 + \cos(\psi(l) + \beta))\,dl \quad (15)$$

Similarly, for the back surface, with array tilt angle equal to $180° - \beta$, the total power output per unit farm area due to diffuse light ($I_{PV:DNI}^{Farm}$) is the sum of $I_{PV:DNI}^{F,Farm}$ and $I_{PV:DNI}^{B,Farm}$.

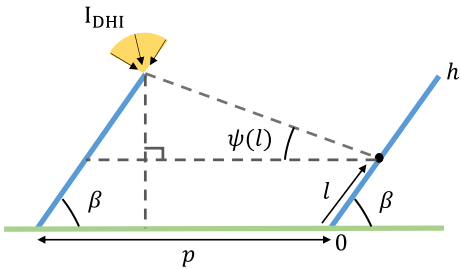

Fig. 2 Schematic diagram for diffuse light collection on the panel

### 2.3.3. Albedo light collection

The estimation of the collection of albedo light is the most complex and lengthy amongst the three components and the complete formulation can be found in appendix section S1. The final expression is as follows:

$$I_{PV:Alb}^{Farm} = I_{PV:Alb.dir}^{Farm} + I_{PV:Alb.diff}^{Farm} \quad (16)$$

where total albedo light collected on the panel ($I_{PV:Alb}^{Farm}$) is the sum of albedo due to direct light ($I_{PV:Alb.dir}^{Farm}$) and diffuse light ($I_{PV:Alb.diff}^{Farm}$).

### 2.3.4. Energy Output

Finally, the total power generated due to light collected from all the components of irradiance is the combination of Eqs. (9), (12) and (20), as follows

$$I_{PV}^{Total} = I_{PV:DNI}^{Farm} + I_{PV:DHI}^{Farm} + I_{PV:Alb}^{Farm} \quad (17)$$

For energy output, we integrate the power generated over the desired period of time. We define the energy yield per pitch of a farm over a period of one year as yearly yield ($YY$).

$$YY(p, \beta, h, \gamma, R_A) = \int_0^1 I_{PV}^{Total}(p, \beta, h, \gamma, R_A)\,dY \quad (18)$$

Next, we will consider the overall optimization of $LCOE^*$.

## 3. Optimization Methodology

Once we have the energy output ($YY$) from a farm in terms of the fundamental variables ($p, \beta, h, \gamma, and\ R_A$), we incorporate this information into Eq. (6). For example, let us assume that albedo ($R_A = 0.5$), array azimuth ($\gamma = 180°\ or\ 0°$), module height ($h = 1\ m$), and $M_L$ is fixed. Therefore, the optimization of the bifacial farm reduces to the optimization for 2 physical parameters ($p, \beta$) to minimize $LCOE^*$. Fortunately, $p$ and $\beta$ are correlated, as explained in the next section.

### 3.1. Mutual shading constraint correlating $p$ and $\beta$

Mutual shading refers to the shading (or obstruction of direct sunlight) of one array of panels by the neighboring array. Partial shading reduces light collection and energy output, and the non-uniform illumination increases self-heating and module degradation[27,28,50]. To avoid mutual shading, the arrays are separated by a distance (pitch) equal to the length of the shadow (at 9am in winter) of an array on the ground. The length of the shadow is longest when Sun's elevation is the smallest. This happens on the shortest day of the year, which is 21st December for the northern hemisphere and 21st June for the southern hemisphere. Assuming the farm is regularly active from 9 am, the pitch of the farm is fixed as the length of the longest shadow observed at 9 am on the shortest day. In principle, the turn-on time may be latitude-dependent; however, our presumption of 9 am turn-on standardizes the analysis. The zenith angle of the sun together with the array tilt angle ($\beta$) gives us the pitch equal to the longest shadow at 9 am for a specific location. We calculate the zenith angle (90° − elevation angle) of the sun at 9 am using the Ephemeris model in the Sandia PV library [34]. The analytical formula for pitch is derived using geometry (see Fig. (3)) and expressed below.

$$S_L = h \cos(\gamma - A) \sin\beta \ / \cot\theta_{zw}$$

(19)

$$p_{ns}/h = S_L/h + \cos\beta$$

where $S_L$ is the extended shadow length.

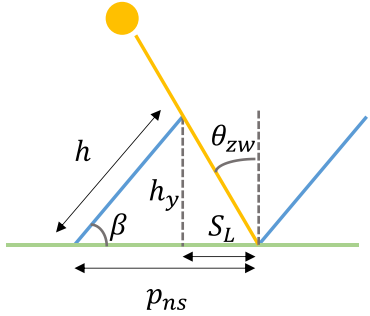

Fig. 3 A two-dimensional schematic of two neighboring arrays separated by a distance $p_{ns}$ for no shading constraint.

Consequently, the mutual shading constraint allows us to calculate the pitch as a function of the array tilt angle, i.e., $p_{ns} = f(\beta)$. The corresponding plot is shown in Fig. S4 in the appendix. Since $p_{ns} = f(\beta)$, and assuming albedo $R_A = 0.5$, the yearly yield $YY$ is now a function of $\beta$ alone, i.e. $YY(p, \beta; h = 1\,m, \gamma = 0°, R_A = 0.5) \rightarrow YY(\beta; h = 1\,m, \gamma = 0°, R_A = 0.5)$. Therefore, for a specific value of $M_L$, $LEOE$ (Eq. 6) is optimized by the optimization of the array tilt angle, namely $\beta_{opt}(M_L)$. Consequently, we vary $\beta$ from 0° to 90° and calculate the yearly energy output of the farm ($YY$) via simulation of the above-mentioned irradiance and light collection models. Further, we calculate the $LCOE^*$ for a particular value of $M_L$, and then find the minima of $LCOE^*$, and the corresponding optimum array tilt angle, $\beta_{opt}$. With $p_{ns}^{opt} = f(\beta_{opt})$, the minimum essential levelized cost of energy ($LCOE_{min}^*$) is

$$LCOE_{min}^* = \frac{p_{ns}^{opt}/h + M_L}{YY(\beta_{opt})}$$

(20)

The simulations yield interesting and noteworthy results which are discussed in the following section.

## 4. Results and Discussion

As noted previously, the $LCOE^*$ is dependent on several physical and economic parameters. Assuming $h = 1\,m, \gamma = 0°\ or\ 180°$, and $R_A = 0.5$, and finding a relationship between pitch and array tilt angle, $p = f(\beta)$, we finally derived Eq. (20) that depends only on two variables, namely Cost ratio ($M_L$) and array tilt angle ($\beta$). In the following discussions, we will calculate $M_L$-dependent optimum tilt-angle for a specific location (Sec. 4.1) and globally (Sec. 4.2), optimum-tilt angle dependent yearly yield (Sec. 4.2.2), $M_L$-dependent minimum levelized cost of energy (Sec. 4.2.3), and $LCOE^*$-reduction of bifacial design over monofacial (Sec. 4.2.4). We also analyze results for three special cases of $M_L$ over various $\beta$ to understand the location-specific design optimization of farms.

### 4.1. Tilt-optimized Bifacial Solar Farm in Washington, D.C.

Before we proceed with the global analysis, it is instructive to examine the optimization of $\beta$ to minimize the $LCOE^*$ for a specific location, e.g., Washington DC (38.91°N, 77.04°W). The variation of $LCOE^*$ with the array tilt angle is analyzed and plotted for two extreme cases shown below.

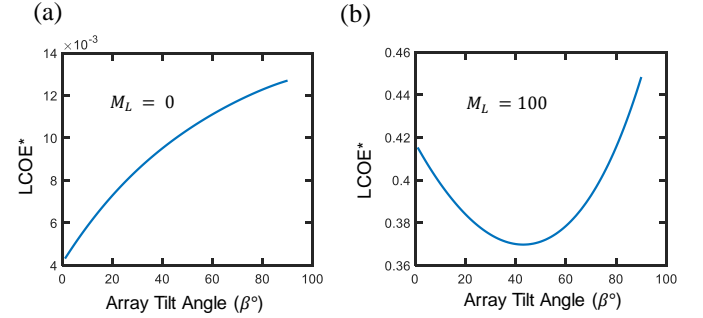

Fig. 4 $LCOE^*$ vs. $\beta$ (a) $M_L = 0$, (b) $M_L = 100$.

**Case 1: $M_L \rightarrow 0$ (i.e., $c_M \ll c_L$, the land is much more expensive than the modules)**

In this case, we observe from Fig. 4(a) that the minimum $LCOE^*$ occurs at an optimum tilt angle, $\beta_{opt} = 0°$. This implies that wherever the land is much more expensive than the module, the installed panels should be installed flat on the ground, stacked end-to-end next to each other, to ensure maximum packing of the arrays (i.e., period ($p$) = height ($h$)). Moreover, this case is applicable to most land-constrained designs i.e., megacity installations where land is limited.

**Case 2: $M_L = 100$ (i.e., $c_M \ll c_L$ the modules are much more expensive than the land)**

For the other extreme case, where the modules are much more expensive than the land, Fig. 4(b) shows the variation of $LCOE^*$ with array tilt angle for the same location (38.91°N, 77.04°W). Intriguingly, we find that there exists an optimum $\beta$ (~42°) to achieve minima in $LCOE^*$. The optimum angle for bifacial farm is higher (~10°) compared to angle-optimized monofacial cells (i.e., $\beta$~32°).

**Case 3: $0 < M_L < 100$ ($c_M \sim c_L$)**

Several locations in the world might fall in this category, where the land and module costs are comparable. Historically, $M_L$ has ranged from ~0.1 to 15. Thus, we specifically explore the variation of $LCOE^*$ and $\beta_{opt}$ over this typical range of $M_L$ (see inset of Fig. 5(a),(b)).

For Washington, DC (38.91°N, 77.04°W), we find that the minimum $LCOE^*$ increases linearly until $M_L \sim 8$ and then deviates from linearity (increases sub-linearly) for higher $M_L$ (see Fig. 5(a)). This slope-change is understood by using Fig. 5(b) where the optimum tilt angle ($\beta_{opt}$) increases abruptly at $M_L^* \sim 8$. A similar trend is seen in $YY$ vs. $M_L$. For $M_L < 8$, the

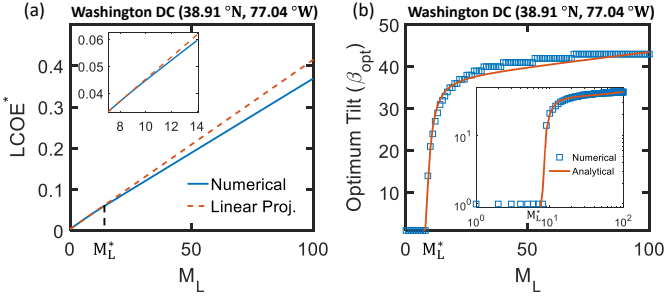

Fig. 5 (a) $LCOE^*$ vs. $M_L$. (b) Comparison between numerical analytical values of $\beta_{opt}$ vs. $M_L$. Inset of (a) and (b) show exactly where the transition occurs ($M_L^* \sim 8$).

higher yearly yield associated with the tilted modules is negated by the high land cost (low $M_L$).

Interestingly, the optimum tilt angle (Fig. 5(b)) can be approximated in terms of $M_L$ using the following empirical formula.

$$\beta_{opt}^{ana} = c_1 M_L + \exp\left(c_2\left(1 - \frac{1}{(M_L - M_L^*)^{c_3}}\right)\right) \quad (21)$$

where $c_1 = 0.07$, $c_2 = 3.6$ and $c_3 = 1.6$ are location-based empirical constants. Here, $M_L^* \sim 8$ is the threshold cost ratio, below which $\beta_{opt} = 0°$.

Note that since $YY$ depends on latitude/longitude of a location, so does the $M_L^*$ for that location. The location-specific $M_L^*$ (typically 0-10) is calculated numerically (see Fig. S5 in appendix).

### 4.1.1. Discussion of result from Washington, D.C.

To rationalize the trends in $LCOE^*$, we examine the (relative) contributions of essential costs vs. energy yield. In this regard, we reconsider Eq. (6), where total cost $C = p/h + M_L$ and yearly energy yield is $YY$. Hence, the relative error in $LCOE^*$ is given by the following expression.

$$\frac{\Delta LCOE^*}{LCOE^*} \equiv \frac{\Delta C(p/h, M_L)}{C(p/h, M_L)} - \frac{\Delta YY(p/h, \beta)}{YY(p/h, \beta)} \quad (22)$$

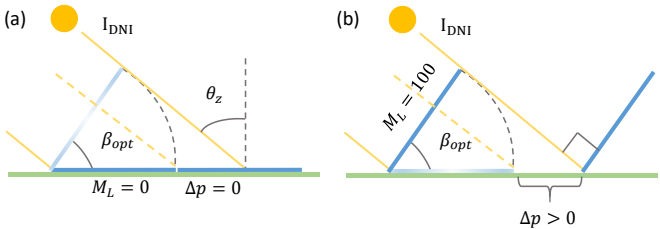

Fig. 6 (a) For $M_L = 0$, stacking panels next to each-other (i.e., $\Delta p = 0$) maximizes light collection per unit area of the module. The cosine incidence of direct light on the panels is shown. (b) For $M_L = 100$, the light collection is increased with titled and optimally-separated modules.

Here, the cost is a function of $p/h$ and $M_L$ whereas yearly energy yield is a function of $p/h$ and $\beta$. With the objective of achieving negative (or less positive) $\Delta LCOE^*$, let us now examine the results through the above equation, and Fig. 6.

For $M_L = 0$, the land cost is much higher than the module cost, which necessitates maximum collection of light for the lowest land area (pitch). This is achieved when $\Delta p \to 0$ (as shown in Fig. 6(a)), such that the panels collect all the light falling on the ground ($I_{GHI}$) with least amount of land used. Although $I_{GHI}$ is collected completely by the farm, each of the panels collect cosine of direct light ($I_{DNI}\cos(\theta_Z)$, see Fig. 6(a)) throughout the day. Fig. 6 shows that $\Delta p \to 0$ reduces the total light collection as compared to $\Delta p > 0$ (i.e., $\Delta YY < 0$), but this is counterbalanced by the reduction in the cost of land ($\Delta C < 0$). Overall, $LCOE^*$ is reduced (i.e., $\Delta LCOE^* < 0$). Therefore, in this case, the optimum design involves stacking the panels next to each other flat on the ground to minimize $LCOE^*$.

For $M_L = 100$, the land cost is much lower than the module cost, and it is vital to maximally collect light per unit area of the module. This is attained by collecting direct light incident perpendicular onto the front face of the panel and collecting the albedo light on both faces to increase the energy yield i.e., $\Delta YY > 0$. This comes at the expense of increased cost ($M_L = 100$) i.e., $\Delta C > 0$, but overall the magnitude of $\Delta YY$ is greater than that of $\Delta C$, therefore, $LCOE^*$ decreases (i.e., $\Delta LCOE^* < 0$). Fig. 6(b) shows the schematic of the economically viable design where $I_{DNI}$ falls perpendicular to the front face and $\Delta p$ depends on the mutual shading constraint at that location.

We now understand the variations in $LCOE^*$ and optimum tilt angle for various cases at a particular location (Washington, D.C.). This enables us to scrutinize the global trends in optimum tilt angle ($\beta_{opt}$), yearly energy yield ($YY$), and $LCOE^*$.

### 4.2. Global Analysis for Tilt-optimized Bifacial Farms

In this section, we will examine the results of worldwide simulations and draw inferences from these results. Note that beyond |latitude| $> 60°$, the days are extremely short during winters, hence, the simulation yields unphysical values for $p_{ns} = f(\beta)$. Therefore, we do not include the results for |latitude| $> 60°$.

### 4.2.1. Optimum tilt angle ($\beta_{opt}$)

For Case 1 where module cost is much smaller than the land cost ($M_L \to 0$), Fig. 7(a) clearly shows that the panels should be deployed horizontally on the ground. Similar to Washington, D.C., this conclusion unequivocally holds for all the locations around the world. The explanation also remains the same as described earlier in Sec. 4.1.1.

For Case 2 where modules are much more expensive than land ($M_L = 100$), Fig. 7(c) illustrates the latitude-dependent optimum tilt angle for a cost-optimized solar farm, the physical explanation of which was explained in detail in Sec. 4.1.1. The latitude-wise increasing trend in optimum angles is broken for places such as Central and East China (Chongqing and Xi'an) due to the contributions from diffused light.

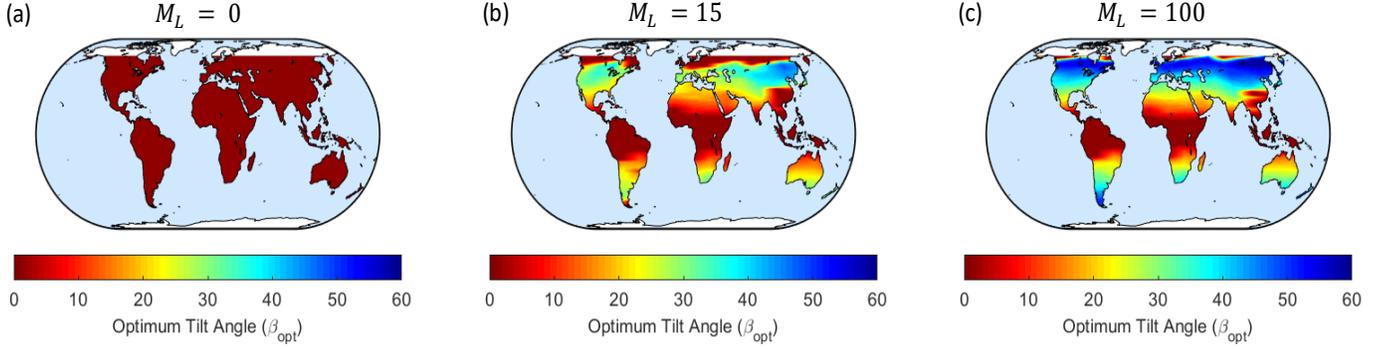

Fig. 7 Global map of optimum array tilt angle associated with the minimum LCOE* for that location: (a) $M_L = 0$ (b) $M_L = 15$ (c) $M_L = 100$

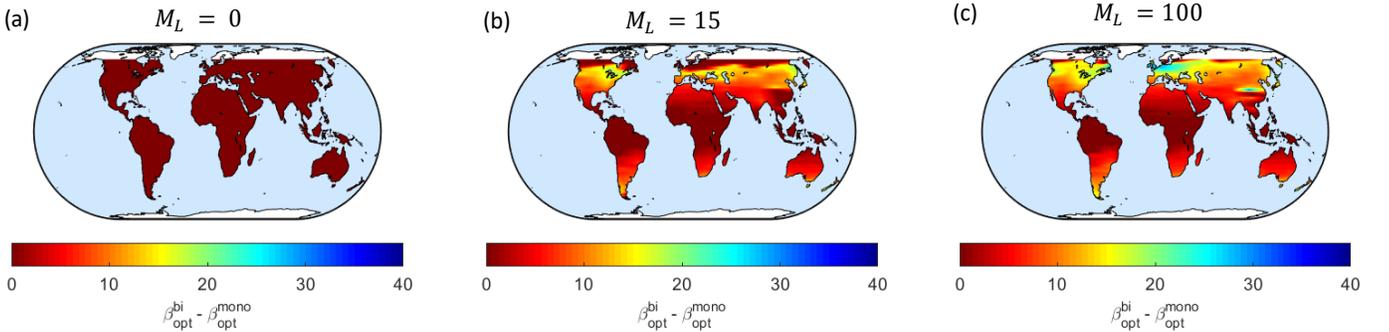

Fig. 8 Global map of difference in optimum array tilt angle between bifacial and monofacial designs (a) $M_L = 0$ (b) $M_L = 15$ (c) $M_L = 100$

For Case 3 with comparable module vs. land-cost, typical values of $M_L(\sim 15)$, Fig. 7(b) displays the optimum tilt angle between Case 1 and 2, in accordance with the plot in Fig. 4(b). Note that, at higher latitudes, as the optimum tilt angle increases (for Cases 2 and 3), the East-West facing vertical bifacial farms become cost competitive with tilted bifacial farms (although with an increased land area) [25,28,31,51].

Fig. 8 demonstrates the difference in the optimum tilt angle for bifacial vs. monofacial design. For $M_L > 0$, we see $\beta_{opt}$ for bifacial is greater than that of monofacial farms in order take advantage of the bifaciality and collect more light at the back surface of the panel. $\Delta \beta_{opt} \sim 15° - 20°$ for $|\text{latitude}| > 30°$ and areas with high diffuse light fraction, e.g. Canada, Western Europe, Central China etc. An added benefit to higher tilt angle is a reduction in cleaning cost due to a decrease in soiling [52].

### 4.2.2. Local and Global Yearly Energy Yield (YY)

Next, we analyze the yearly energy output from a tilt-optimized solar farm for two limiting cases: (i) $M_L = 0$, i.e., the module costs negligible as compared to land. These locations can be cosmopolitans, megacities, and cities, and, in general, densely populated areas; (ii) $M_L = 100$ where modules are much more expensive compared to land. These can be remote locations and sparsely populated areas. Fig. S3 (in appendix) shows the global map for these two cases. We notice a slight improvement in the yearly yield for $M_L = 100$ as compared to $M_L = 0$. This is also evident from the simulation for a single location (e.g., Washington DC) where YY for $M_L = 0$ case is $231\ kWh/m^2$ while YY for $M_L = 100$ is $278\ kWh/m^2$. The is because the collection of direct light and albedo light are enhanced for $M_L = 100$ as compared to $M_L = 0$ (both for the same city). We also find that the energy yield steadily reduces with increasing latitude, except for some specific places like Central and East China, and West Brazil where the trend is broken. These regions are characterized by lower clearness index with higher fraction of diffuse light compared to places near the equator.

### 4.2.3. Minimum Levelized Cost of Energy (LCOE*min)

Given the information about optimum tilt-angle (Sec. 4.1, 4.2.1) and energy yield (Sec. 4.2.2), we can now calculate minimum $LCOE^*$ for locations around the world. Fig. 9 shows that, unlike the yearly energy yield world maps, the $LCOE^*_{min}$ for $M_L = 0$ follows a similar (but inverse) trend as YY for $M_L = 0$, i.e., the maxima in YY are exactly the minima in $LCOE^*_{min}$. On the other hand, these two parameters for $M_L = 100$ do not show an exact inverse trend. This difference elucidates the role of essential costs in $LCOE^*$ calculations. world maps depend sensitively on module vs. land costs.

When the land cost is very high as compared to module cost ($M_L = 0$), then the $LCOE^*$ is dominated by the yearly energy yield, and hence we observe the similar (but inverse) trend since $LCOE^*$ is inversely proportional to $YY$.

Whereas in the reverse case ($M_L = 100$), $LCOE^*$ is dominated by $M_L$. Therefore, we see in Fig. 9(c) that $LCOE^*$

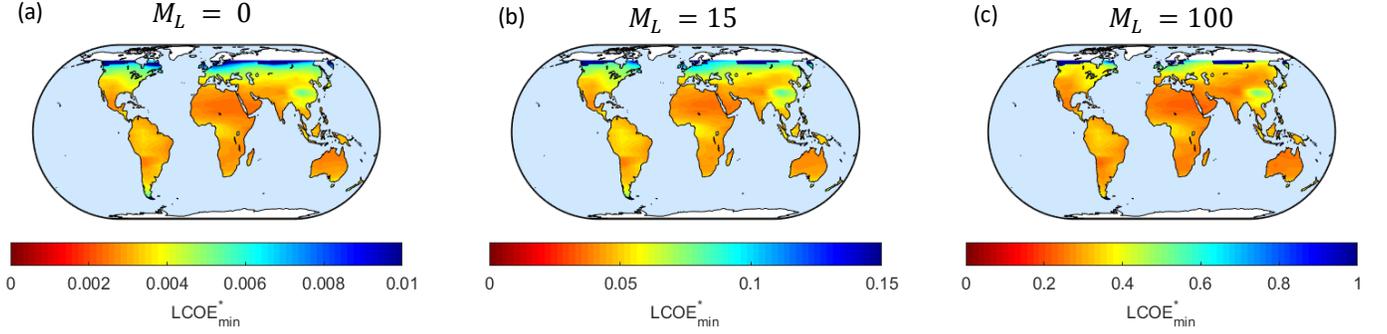

Fig. 9 $LCOE^*_{min}$ global maps for (a) $M_L = 0$ (b) $M_L = 15$ (c) $M_L = 100$

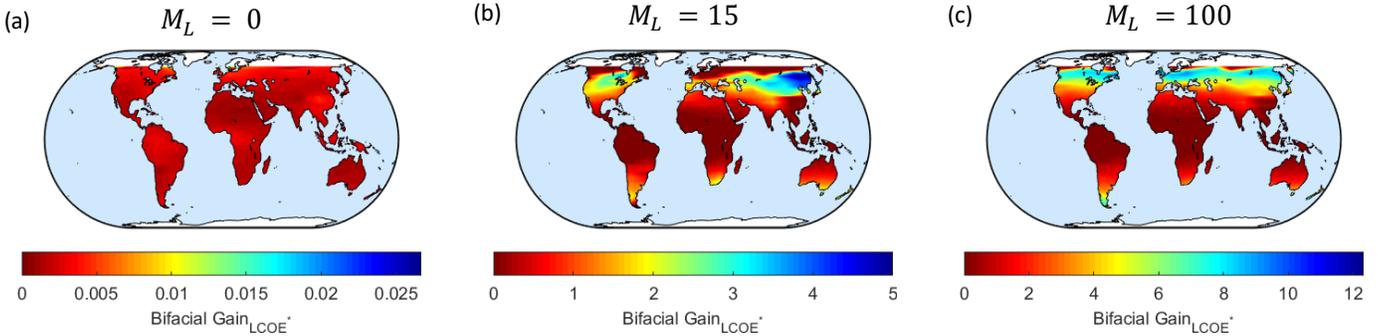

Fig. 10 Bifacial gain in terms of $LCOE^*$ reduction for (a) $M_L = 0$ (b) $M_L = 100$

remains almost uniform throughout the globe pertaining to the much larger value of $M_L$ as compared to $p/h$.

### 4.2.4. LCOE* improvement of bifacial over monofacial

To scrutinize the viability of bifacial solar farms, it is important to juxtapose the performance of bifacial solar farms with their monofacial counterparts. The cost of bifacial modules has been falling consistently over the years [14,24]. Even though our approach can easily compare farms with different $M_L$, the following discussion assumes an optimistic case where the costs of bifacial and monofacial modules (i.e. $M_L$) are equal. Fig. 10 shows the absolute percentage decrease in $LCOE^*_{min}$ when we decide to install a bifacial farm vs. a monofacial farm. Clearly, for $M_L = 0$, bifacial $LCOE^*$ is equal to monofacial $LCOE^*$ since the panels are flat on the ground and the back surface does not collect any light. This result would be different and in favor of bifacial solar farms if the panels are elevated above the ground (and spaced apart) so that the albedo light can be collected. For $M_L = 100$, on the other hand, there is a negligible reduction in $LCOE^*$ for latitudes less than +/- 30° from the equator.

We find that $LCOE^*_{min}$ reduces by ~6% for |latitude| > 30°. Since the panels are tilted facing the direct light for $M_L = 100$, hence this observation indicates that a bifacial farm is more economical than a monofacial farm in terms of $LCOE^*$, for places where diffuse light is a dominant component of light over direct light. The analysis in Sec. 4.1.1 justifies this conclusion.

Finally, due to the improvement in energy yield for bifacial solar farms for these locations, there is a decrease in $LCOE^*$ (~ 2 − 8 %) for a bifacial farm compared to a monofacial farm, which is displayed clearly in Fig. 10(c) (e.g. Northern USA, Germany, UK, and Central Asia). We have assumed that the monofacial and bifacial costs are similar to produce a Bifacial Gain$_{LCOE^*}$ of 2 − 8%. In other words, bifacial modules should be no more than 2 − 8% expensive for the solar farm to be economically viable [53]. Additional margin in LCOE may be obtained when the modules are elevated, and the reduced cleaning cost associated with higher tilt angle is accounted for. These would be important topics of future research.

## 5. SUMMARY AND CONCLUSIONS

We have parametrically explored the economic viability of ground-mounted tilted bifacial solar farms and explained how the farm topology must be optimized for a given location, and module and land cost considerations. We have redefined the levelized cost of energy ($LCOE$) in terms of 'essential levelized cost of energy' ($LCOE^*$) which is ultimately a function of module to land cost ratio ($M_L$) and array tilt angle ($\beta$). The redefined $LCOE^*$ decouples cost analysis from energy yield modeling, thereby dramatically simplifying the optimization of solar farms based on new technologies.

Using a previously developed global irradiance model [28], we calculated the spatial distribution of light on the ground and panel faces while considering all variations of shadows for all the locations in the world. The collection of direct, diffuse, and albedo light on the panels were then integrated over time to obtain the yearly yield for the specific solar farm configuration (defined by panel tilt and array period). Once we correlate the configuration of a farm to the cost of its installation and the yearly yield, we can determine $LCOE^*$.

The panel tilt $\beta$ defines the array period, which in turn sets the number of panels required in a solar farm. Therefore $\beta$ is implicitly related to the farm cost (and of course the energy). In the end, cost ratio ($M_L$) and array tilt angle ($\beta$) are the handles to control the $LCOE^*$. For a fixed $M_L$, we numerically and analytically found an optimum tilt angle ($\beta_{opt}$) for each location.

Our analysis leads to the following key **conclusions**:

- For places where land is scarce and expensive ($M_L \to 0$), panels should be laid flat on the ground ($\beta_{opt} = 0°$) to ensure maximum energy collection over a given land area. On the contrary, for practical values of $M_L (\sim 1 - 15)$ when the land is relatively inexpensive, panels have location specific optimum tilt ($\beta_{opt} \sim 0° - 60°$) to achieve least $LCOE^*$.

- PV installers can use an analytical expression of the form of Eq. (21) to find the location-specific optimum array tilt angle ($\beta_{opt}$) as a function of $M_L$. Moreover, $\beta_{opt}$ is constantly zero until a threshold value ($M_L^*$) of cost ratio which varies with the location (latitude/longitude).

- The difference in optimum tilt angle ($\Delta\beta_{opt}$) between bifacial and monofacial designs can reach up to $15° - 20°$ for |latitude| $> 30°$ and places with high diffuse light fraction, e.g., Canada, Western Europe, Central China, etc. Moreover, higher tilt angle makes the design soiling-resistant, in turn reducing cleaning cost.

- For the same module-to-land cost ratio and similar lifetimes (reliability), ground-mounted bifacial solar farm design is more economically viable over monofacial design for locations where the diffuse fraction is high. The relative reduction in $LCOE^*$ (Bifacial Gain$_{LCOE^*}$) is $\sim 2 - 8\%$, for bifacial solar farm design over monofacial for locations with higher fractions of diffuse light (low clearness index, $k_T$) e.g., locations with |latitude| $> 30°$ (Central Europe, Northern parts of North America, and Central China). Alternatively, bifacial modules can be at most $\sim 2 - 8\%$ more expensive compared to monofacial modules for a bifacial solar farm to be cost-competitive compared to a monofacial farm.

Although this is the first report of LCOE-optimized farm design, the present work can be generalized in a number of ways. One can use the current approach to easily account for location-specific albedo and tilt-angle. Furthermore, currently the bifacial panels are slightly more expensive (but also known to be more reliable [12,19]) than monofacial modules. This cost and reliability differences can be accounted for easily in our formulation. Instead of ground-mounted panels, the farm design can deploy elevated panels. The elevation of panels could increase the albedo light collection depending on the farm design, but this gain must be balanced against the increase in the installation (module) cost. It is also possible to sculpt the ground to increase albedo and re-optimize the configuration.

To conclude, the reduction in $LCOE^*$ through optimized farm design and continually reducing bifacial module prices makes bifacial PV technology an economically preferable alternative over monofacial solar farm.

### ACKNOWLEDGMENT

The authors gratefully acknowledge the discussions with Drs. Peter Bermel (Purdue), Chris Deline (National Renewable Energy Laboratory)), and Josh Stein (Sandia National Laboratory).

### APPENDIX

#### S1: Albedo Light Collection
For the albedo light collection, we will start with the same approach used by Khan et al. [ref.] for vertical bifacial panels and generalize it for the tilted bifacial design.

We first account for albedo from direct light. The direct light collected on the ground is given by:

$$I_{Gnd:DNI}(x) = I_{DNI} \cos\theta_Z(t) \tag{23}$$

where $s_1(t) < x < p - s_2(t)$. Here $s_1$ is the time-varying shadow length in the morning and $s_2$ is the time-varying shadow length in the afternoon (see Fig. 5). $s_1$ and $s_2$ can be calculated using Eq. (7) on $S_L(t)$. The unshaded part $p - s_1$ on the ground subtends angle $(\psi_0, \pi/2)$ at the point $l$ on the front panel face (see Fig. 5) in the morning while the unshaded part $p - s_2$ subtends $(\psi_1, \psi_2)$ in the afternoon. These angles are geometrically calculated as:

$$\left. \begin{array}{l} \psi_0(l) = \tan^{-1}\left[\dfrac{-(p - s_1(t) + l\cos\beta)}{l\sin\beta}\right] - \beta \\[2mm] \psi_1(l) = \tan^{-1}\left[\dfrac{-(p + l\cos\beta)}{l\sin\beta}\right] - \beta \\[2mm] \psi_2(l) = \tan^{-1}\left[\dfrac{-(s_2(t) + l\cos\beta)}{l\sin\beta}\right] - \beta \end{array} \right\} \tag{24}$$

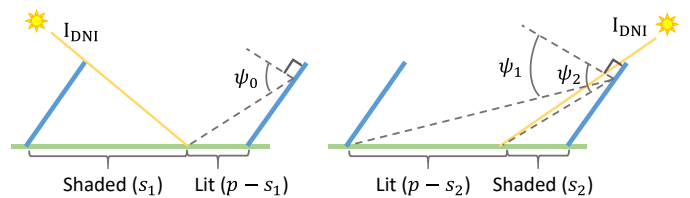

Fig. S1 Schematic for albedo from direct light: (a) before noon and (b) afternoon

Using these angles, we find the view factor from the point $l$ from the ground to the unshaded part of the ground.

$$F_{dl-gnd} = \frac{1}{2}\left(1 - \sin\psi_0(l)\right), \text{ in morning;}$$

$$= \frac{1}{2}\left(1\sin\psi_2(l) - \sin\psi_1(l)\right), \text{ in}$$
afternoon.
$$(25)$$

Therefore, the power generated per area of the panel for the front face is given by

$$I_{PV:Alb:dir}^{F,Panel}(l) = I_{Gnd:DNI} \times R_A \times F_{dl-gnd}(l) \times \eta_{diff} \quad (26)$$

where $R_A$ is the fraction of incident light reflected from the ground. Thus, the power generated per pitch of the farm is given by

$$I_{PV:Alb:dir}^{F,Farm}(l) = 1/h\int_0^h I_{PV:Alb:dir}^{F,Panel}(l)\,dl \quad (27)$$

By incorporating the collection from the back surface, we will have the total albedo due to direct light ($I_{PV:Alb:dir}^{Farm}$).

For albedo from the diffuse light we combine the method used by Khan et al. and the view factors for our tilted panels. Similar to the collection of direct light on the ground, we find the collection of diffuse light on the ground taking into account the masking due to adjacent panels/arrays. Fig. 6(a) illustrates the masking and collection of diffuse light at a point $x$ positioned between two adjacent panels. The angles $[\theta_1, \theta_2]$ subtended at point $x$ by the topmost points of the panels are given by

$$\left.\begin{array}{l} \theta_1(x) = \pi - \tan^{-1}\left[\dfrac{h\sin\beta}{h\cos\beta - x}\right] \\[2mm] \theta_1(x) = \tan^{-1}\left[\dfrac{h\sin\beta)}{p - x + h\cos\beta}\right] \end{array}\right\} \quad (28)$$

The view factor from the ground to the sky is $F_{dx-sky}(x) = 1/2[\sin(\pi/2 - \theta_2) - \sin(\theta_1 - \pi/2)]$, which can be used to find $I_{Gnd:DHI}$ as

$$I_{Gnd:DHI} = 1/p\int_0^p I_{DHI} \times F_{dl-gnd}(x)\,dx \quad (29)$$

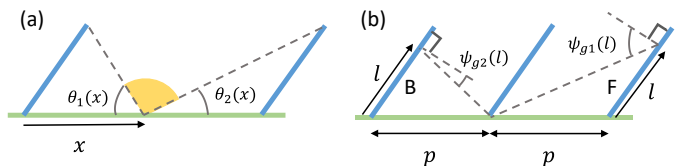

Fig. S2 Schematic for (a) diffuse light collection on the ground (b) view factor calculation for albedo light collection

Now, the albedo light collected on the panels from the ground requires us to calculate the view factor from the ground to the panels (see Fig. 6(b)). This view factor is given by $F_{dx-dl-F}(x) = 1/2[1 - \sin(\psi_{g1}/2)]$ for front face, and $F_{dx-dl-B}(x) = 1/2[1 - \sin(\psi_{g2}/2)]$, where

$$\left.\begin{array}{l} \psi_{g1} = \dfrac{\pi}{2} - \beta - \tan^{-1}\left[\dfrac{l\sin\beta}{p + l\cos\beta}\right] \\[2mm] \psi_{g2} = \dfrac{\pi}{2} - \beta - \tan^{-1}\left[\dfrac{l\sin\beta}{p - l\cos\beta}\right] \end{array}\right\} \quad (30)$$

Finally, we arrive at the power per unit panel area as follows

$$I_{PV:Alb:diff}^{F,Panel}(l) = I_{Gnd:DHI} \times R_A \times F_{dl-gnd}(l) \times \eta_{diff} \quad (31)$$

and the power per pitch of farm is

$$I_{PV:Alb:diff}^{F,Farm} = 1/h\int_0^h I_{PV:Alb:diff}^{F,Panel}(l)\,dl \quad (32)$$

Adding the light collected at the back surface, we get the total albedo due to diffuse light ($I_{PV:Alb:diff}^{Farm}$).

Hence, total albedo light collection is given by

$$I_{PV:Alb}^{Farm} = I_{PV:Alb:dir}^{Farm} + I_{PV:Alb:diff}^{Farm} \quad (33)$$

### S2: Mutual Shading Constraint

The mutual shading constraint at a particular locations on the shortest of day of the year (21st December for Northern Hemisphere and 21st June for the Southern Hemisphere) gives us a relation between the pitch (period) of the farm and tilt angle of the panels (arrays). Fig. S3 shows this relationship.

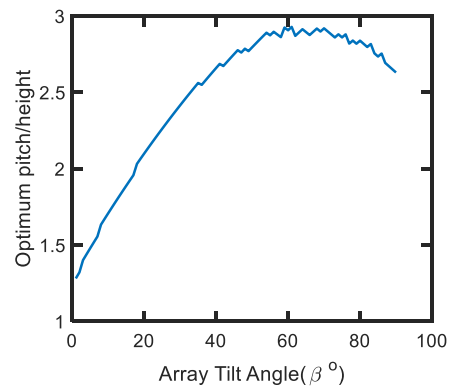

Fig. S3 The plot shows the pitch as a function of array tilt angle achieved using the mutual shading constraint on the shortest day of the year for Washington, D.C.

### S3: Yearly Yield

The yearly energy yield for the two extreme cases is shown in Fig. S3. Notice the high energy yield at Sahara Desert and relatively low yields at places with high fraction of diffuse light e.g., Central China and places with |latitude| > 30°. The trends of $LCOE^*$ and $YY$ are perfectly inverse of each other for $M_L = 0$ while $M_L = 100$ shows the effect of high cost ratio leading to weaker dependence of $LCOE^*$ on $YY$.

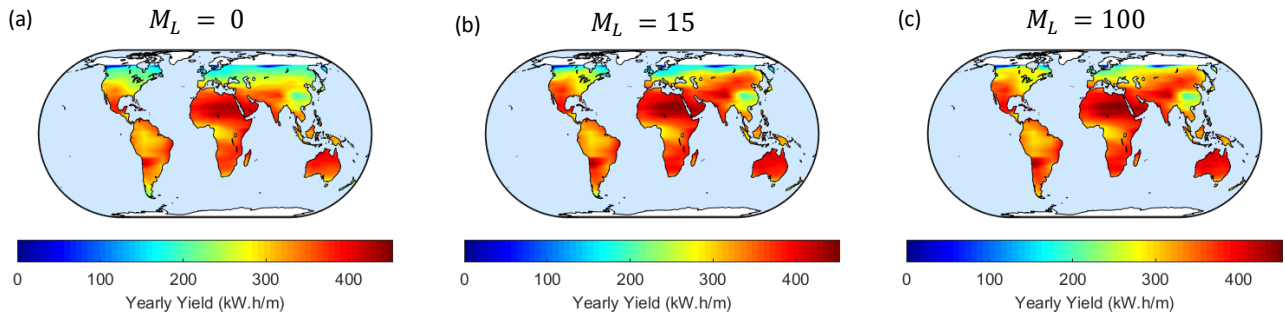

Fig. S4 Global maps showing yearly energy yield of bifacial solar farms for (a) $M_L = 0$ (b) $M_L = 100$.

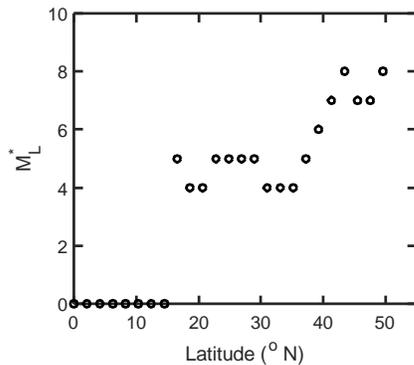

Fig. S5 Typical values of $M_L^*$ with increasing latitude at longitude = $0°$.